\documentclass{osa-article}

\journal{opticajournal}

\articletype{Research Article}

\begin{document}

\title{Ultrashort laser pulses with chromatic astigmatism}

\author{Spencer W. Jolly}

\address{Service OPERA-Photonique, Université libre de Bruxelles (ULB), Brussels, Belgium}

\email{spencer.jolly@ulb.be}

\begin{abstract}
Ultrashort laser pulses are described having chromatic astigmatism, where the astigmatic phase varies linearly with the offset from the central frequency. Such a spatio-temporal coupling not only induces interesting space-frequency and space-time effects, but it removes cylindrical symmetry. We analyze the quantitative effects on the spatio-temporal pulse structure on the collimated beam and as it propagates through a focus, with both the fundamental Gaussian beam and Laguerre-Gaussian beams. Chromatic astigmatism is a new type of spatio-temporal coupling towards arbitrary higher complexity beams that still have a simple description, and may be applied to imaging, metrology, or ultrafast light-matter interaction.
\end{abstract}

\section{Introduction}
\label{sec:intro}

Chromatic aberrations in the context of ultrashort electromagnetic pulse-beams can be equivalently viewed as spatio-temporal couplings (STCs)~\cite{akturk10}. These are generally thought of as detrimental to laser-matter interaction experiments because they cause an increase in the focused pulse duration and therefore a decrease in the intensity~\cite{bourassin-bouchet11}. However, especially recently, STCs are beginning to be thought of as an avenue for control as well. In the realm of laser-based acceleration for example, there are various schemes where despite having a lower peak intensity, space-time shaped beams are predicted to produce a higher net acceleration~\cite{debus19,jolly19-1,palastro20,caizergues20,jolly20-2}.

The link between spatio-spectral and spatio-temporal aberrations is important, because the properties of many optical components and systems are best characterized by their spectral properties (dispersion, etc.). Prisms have angular dispersion which is linked to pulse-front tilt in an ultrashort pulse~\cite{bor93}, and singlet lenses have a frequency-varying focal length which leads to pulse-front curvature~\cite{bor88,bor89-1}. Diffractive optical lenses are also well-known to have inherently strong chromatic focusing, which can lead to significant reshaping of an ultrashort pulse around it's focus~\cite{alonso18}. Indeed, for diffractive optical elements and meta-optics in general this chromatism is often seen as a disadvantage and a design challenge, but the opportunity for control remains.

In this work we will describe and model ultrashort pulses that have chromatic astigmatism, where there is a frequency-varying spatial phase (wavefront) aberration that no longer has the simple symmetry of the case of a prism or a simple lens---i.e. not along one cartesian or cylindrical coordinate. We will describe in the nearfield (collimated) and farfield (focused) spaces, for fundamental Gaussian and Laguerre-Gaussian beams, and with zero or quadratic additional spectral phase. We will often compare this new STC to the known case of pulse-front curvature that is cylindrically symmetric, and this will reveal key characteristics of both situations.

\section{Conceptual description of chromatic astigmatism}
\label{sec:theory}

Ultrashort laser pulse-beams can most simply be described as an electromagnetic beam that has a separable spatial amplitude $p(\vec{r})$ and temporal envelope $g(t)$ such that the electric field for a scalar beam is $E(\vec{r},t)=p(\vec{r})g(t)e^{i(\omega{t}-kz)}$. However, spatio-temporal couplings can make such a description inadequate.

\begin{figure*}[tbh]
	\centering
	\includegraphics[width=\linewidth]{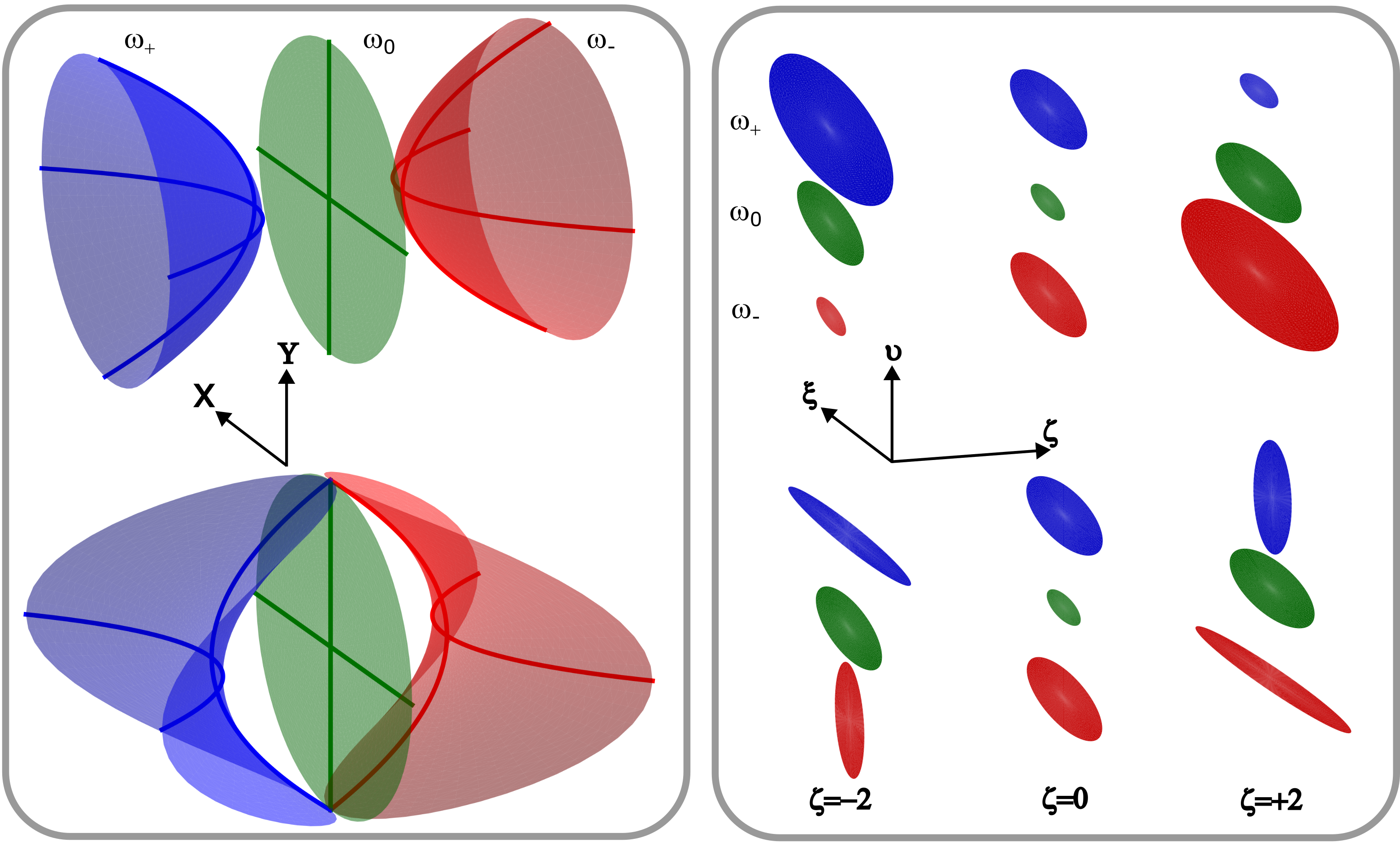}
	\caption{The concept of chromatic astigmatism. Pure chromatic curvature (top) and pure chromatic astigmatism (bottom) are compared. In the nearfield (left) the wavefronts are shown for three frequencies $\omega_-<\omega_0<\omega_+$ (red, green, blue). After focusing to the farfield (right) the beam size is shown for the same three frequencies at three planes in the direction of propagation. \{$\xi,\upsilon,\zeta$\} are the normalized versions of coordinates \{$x,y,z$\} in the focus. Note that the central frequency $\omega_0$ behaves the same everywhere.}
	\label{fig:concept}
\end{figure*}

Chromatic curvature in the nearfield is equivalent to pulse-front curvature (PFC). This can be seen in a pulse whose temporal envelope has a quadratically-varying arrival time with the radius $r$ as $g(t-\alpha|r|^2)$, which can be shown to be equivalent to the radius of curvature $R$ having a frequency dependence such that $1/R(\omega)=2c\alpha\delta\omega/\omega$~\cite{sainte-marie17} (see Fig.~\ref{fig:concept}), where $\delta\omega=\omega-\omega_0$ and $\tau_p=\alpha w_i^2$, where $w_i$ is the nearfield (collimated) beam waist (1/e$^2$ intensity radius). This new temporal envelope depends on both space and time, and results in the electric field no longer being space-time separable.

The above analysis for chromatically-varying curvature or PFC can also be viewed as the defocus Zernike polynomial having a frequency variation proportional to $\tau_p\delta\omega$. Chromatic astigmatism, the main topic of this work, is the extension to the astigmatism Zernike polynomials, which are essentially the next aberrations after defocus in terms of complexity. Vertical astigmatism is the case where the spatial phase is proportional to $x^2-y^2$, in contrast to the defocus term which is proportional to $x^2+y^2=r^2$, so we will consider only this case (see Fig.~\ref{fig:concept}). Keeping generality, we can introduce another set of characteristic parameters \{$\alpha_a,\tau_a$\} ($\tau_a=\alpha_a w_i^2$) that describe the chromatic astigmatism such that the vertical astigmatism Zernike term is proportional to $\tau_a\delta\omega$ and the temporal envelope has a pulse front with a deformed arrival time $g(t-[(\alpha+\alpha_a)x^2+(\alpha-\alpha_a)y^2])$ when there is both PFC and chromatic astigmatism. The asymmetry between the Cartesian coordinates $x$ and $y$ can also be summarized with $\alpha_x=\alpha+\alpha_a$ and $\alpha_y=\alpha-\alpha_a$ (and $\tau_x=\tau_p+\tau_a$ and $\tau_y=\tau_p-\tau_a$).

As can be seen in Fig.~\ref{fig:concept}, there are important links between STCs on a collimated beam and those in the focus. For example, chromatically-varying curvature or PFC on a collimated beam will produce longitudinal chromatism (LC) in the focus, where the different wavelengths are focused to different longitudinal positions. Using the notation of the previous paragraphs, the waist position $z_0$ for each frequency follows the relation: $z_0(\omega)=z_R\tau_p\delta\omega$---higher frequencies are focused later (larger $z$) for a positive $\tau_p$. The same intuition can be made for chromatically varying astigmatism, except that the waist position is different when considering the size in the xz-plane or the xy-plane. Therefore we have $z_{0x}(\omega)=z_R\tau_x\delta\omega$ and $z_{0y}(\omega)=z_R\tau_y\delta\omega$. When there is no chromatically-varying curvature ($\tau_p=0$) and only chromatic astigmatism ($\tau_a\neq0$), $z_{0x}$ and $z_{0y}$ are equal in magnitude and have opposite sign, which is the case shown in Fig.~\ref{fig:concept} and one of the main cases considered in the rest of this work. The following section is essentially a quantitative treatment of the explanations from this section.

\section{Analysis with the fundamental Gaussian beam}
\label{sec:analysis}

We will first focus on the pulse-beam having a Gaussian spatial and temporal profile (the fundamental Gaussian beam), in both the nearfield and farfield.

\subsection{Chromatic astigmatism on the collimated beam}
\label{sec:analysis_NF}

The duality of time and frequency allows for a relatively straightforward representation of chromatic astigmatism on a collimated (nearfield) Gaussian beam. Assuming a beam of width $w_i$ and pulse duration $\tau_0=2/\Delta\omega$, for the complex electric field at a single longitudinal plane $E=Ae^{i\omega_0{t}}$ and $\tilde{E}=\mathcal{F}\{E\}$ we have

\begin{align}
	\tilde{E}(x,y,\omega)&=e^{-(X^2+Y^2)}e^{-i\tau_a\delta\omega(X^2-Y^2)}e^{-\delta\omega^2/\Delta\omega^2} \label{eq:NF_freq} \\
	A(x,y,t)&=e^{-(X^2+Y^2)}e^{-(t-\tau_a(X^2-Y^2))^2/\tau_0^2} \label{eq:NF_time},
\end{align}

\noindent where $X=x/w_i$ and $Y=y/w_i$. One immediately notices that both the frequency and time representations of the electric field have terms where time/frequency and space are unseparable. In the frequency description it is purely on the phase, and in the temporal description it is on the arrival time.

\begin{figure*}[tbh]
	\centering
	\includegraphics[width=\linewidth]{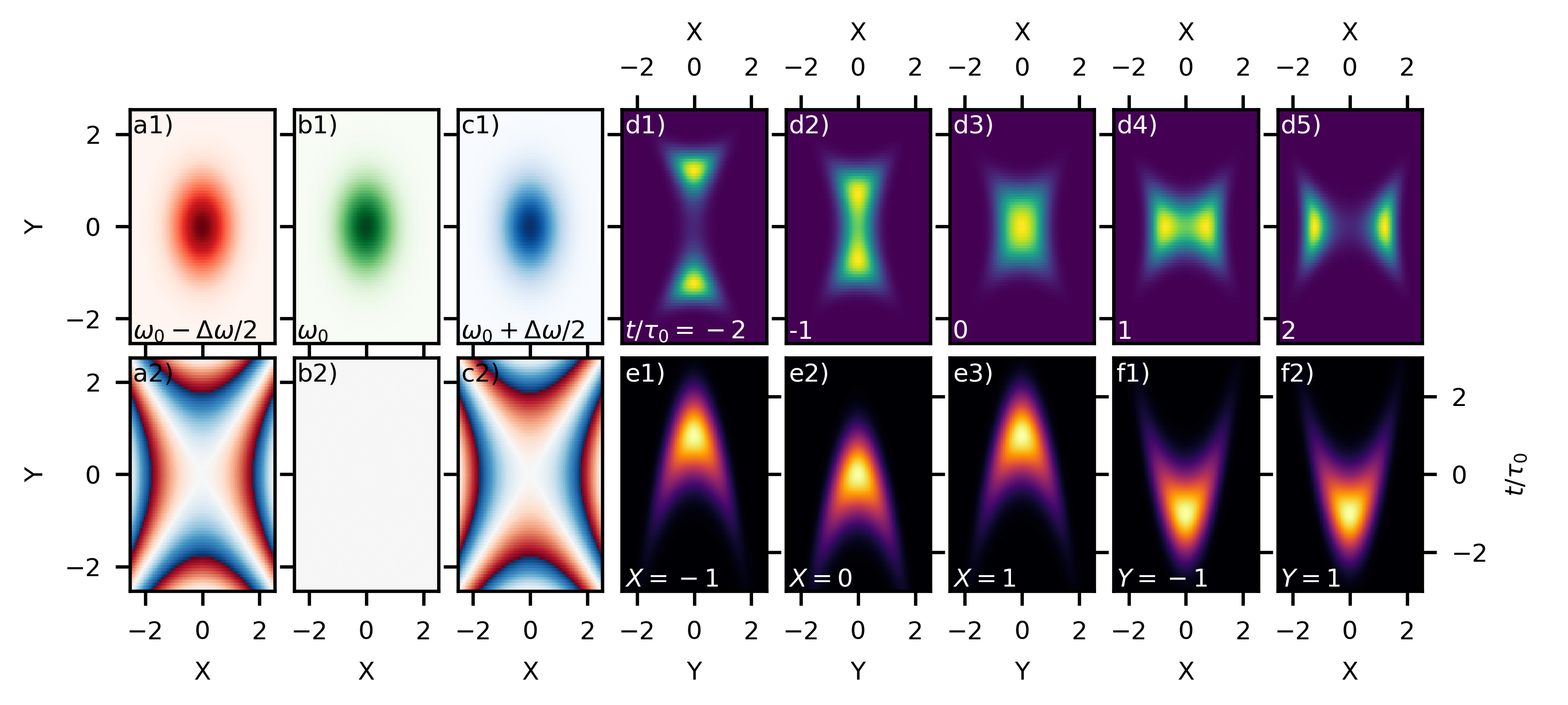}
	\caption{Ultrashort laser pulse with chromatic astigmatism ($\tau_a=\tau_0$) in the nearfield in frequency and time. The different frequency components (a--c) have the same amplitude (a1--c1), but have varying spatial phase. The chromatic astigmatism parametrized by $\tau_a$ and described in Eqs.~\ref{eq:NF_freq}--\ref{eq:NF_time} results in an astigmatic phase (a2--c2) of opposite sign at frequencies above and below $\omega_0$. The pulse in time is characterized by a saddle-like arrivval time, which can be viewed by a time-varying amplitude profile (d), or a spatially-varying temporal profile via slices along X (e) or Y (f).}
	\label{fig:NF}
\end{figure*}

In the frequency space according to Eq.~\ref{eq:NF_freq}, shown in Fig.~\ref{fig:NF}(a--c), the spatial amplitude is still constant across all frequencies. However, the spatial phase is varying with frequency. Indeed, this is the most direct way to view it as chromatic astigmatism---the central frequency has no spatial phase (Fig.~\ref{fig:NF}(b2)), but frequencies above and below the central frequency have astigmatism of equal magnitude but opposite sign (Fig.~\ref{fig:NF}(a2) and (c2)).

Although the concept may be simple, the results in the temporal domain are not necessarily intuitive. The time-integrated spatial amplitude is the same as a fundamental Gaussian without any STCs. However, there is temporal structure. At different times, shown in Fig.~\ref{fig:NF}(d1--d5), the spatial amplitudes are non-trivial. From a different perspective it may be straightforward: slices in Yt-plane for different $X$ values (Fig.~\ref{fig:NF}(e1--e3)) and slices in the Xt-plane at different $Y$ values (Fig.~\ref{fig:NF}(f1--f2)) show the moving local pulse-front curvature that switches orientation depending on the sliced plane. These slices are a clear signature of the saddle shape take by the pulse front in the presence of the chromatic astigmatism seen on inspection of Eq.~\ref{fig:NF}(d--f). The different perspectives in Fig.~\ref{fig:NF}(d--f) are all showing the saddle-shape arrival time of the pulse-front in a different way.

An important additional consideration is the symmetries of the equations. In both frequency-space and in time, the beam is always symmetric with inversion of $x$ and $y$ when accompanied by a certain additional operation in frequency or time, respectively. In frequency space the symmetry is upon reflection over $\omega_0$, and in time it is upon reflection over the origin. More concretely: $\tilde{E}(x',y',\omega')=\tilde{E}(y',x',2\omega_0-\omega')$, and $A(x',y',t')=A(y',x',-t')$. These symmetries can be confirmed in Fig.~\ref{fig:NF}: panels \{a2,c2\} are the same when $X$ and $Y$ are exchanged, and the same is true for \{d1,d5\} and \{d2,d4\}. Panels \{e1,f2\} and \{e3,f1\} are the same when reflected over $t=0$. Accordingly, panels b2 and d3 in Fig.~\ref{fig:NF} are symmetric with an exchange of $X$ and $Y$ (a trivial result) since they are each at the symmetry point in frequency space and time, respectively.

\subsection{Chromatic astigmatism on the focused beam}
\label{sec:analysis_FF}

The electric field of a propagating beam can be written in frequency space as $\tilde{E}=\psi e^{-\delta\omega^2/\Delta\omega^2} e^{-ikz}$ such that $\psi$ is essentially describing the evolving amplitude and phase. The fundamental Gaussian beam in the focus can be described in the most compact way with $\psi_0$ as

\begin{equation}
	\psi_0=fe^{-f\rho^2\omega/\omega_0} \label{eq:FF_Gaussian},
\end{equation}

\noindent where $\rho=r/w_{00}$ and $f=i/(i+\zeta)$, with $\zeta=z/z_R$. Note the term $\omega/\omega_0$ in the exponential, which is necessary to properly describe the time delay developed on the pulse as it diffracts and gains significant curvature outside of the focus. There is the explicit assumption that $z_R$ is frequency-independent (Porras factor $g_0=0$~\cite{porras09}), which requires that $w_{00}$ is the focused beam waist at the central frequency ($w_{00}\equiv{w_0(\omega_0)}$). Indeed, this not only results in the correct time delay due to curvature, but also a frequency-dependent beam waist $w_0\propto1/\sqrt{\omega}$. However, this only has a significant additional effect when the pulse becomes few-cycle. The specific effects of the frequency-dependent beam waist and other Porras factor values are not considered in this work.

When expanded to separate the amplitude and phase components, the compact description produces the commonly known relationships for Gouy phase, phase curvature, beam waist evolution, etc. In the case of $\psi_0$ where there is no STC, one can take the Fourier-transform to calculate the field in time. When there is longitudinal chromatism ($\tau_p\neq0$), $f=i/(i+\zeta-\tau_p\delta\omega)$ and the field in time can no longer be easily calculated except by using numerical integration.

The focused fields of the fundamental Gaussian beam with chromatic astigmatism can be constructed as a relatively simple extension of above, creating $\psi_a$ as follows

\begin{align}
	\psi_a&=\sqrt{f_xf_y}e^{-(f_x\xi^2+f_y\upsilon^2)\omega/\omega_0} \label{eq:FF_CAstig} \\
	f_x&=\frac{i}{i+\zeta-\tau_x\delta\omega} \\
	f_y&=\frac{i}{i+\zeta-\tau_y\delta\omega},
\end{align}

\noindent where $\xi=x/w_{00}$ and $\upsilon=y/w_{00}$, and we are reminded that $\tau_x=\tau_p+\tau_a$ and $\tau_y=\tau_p-\tau_a$.

\begin{figure*}[tbh]
	\centering
	\includegraphics[width=\linewidth]{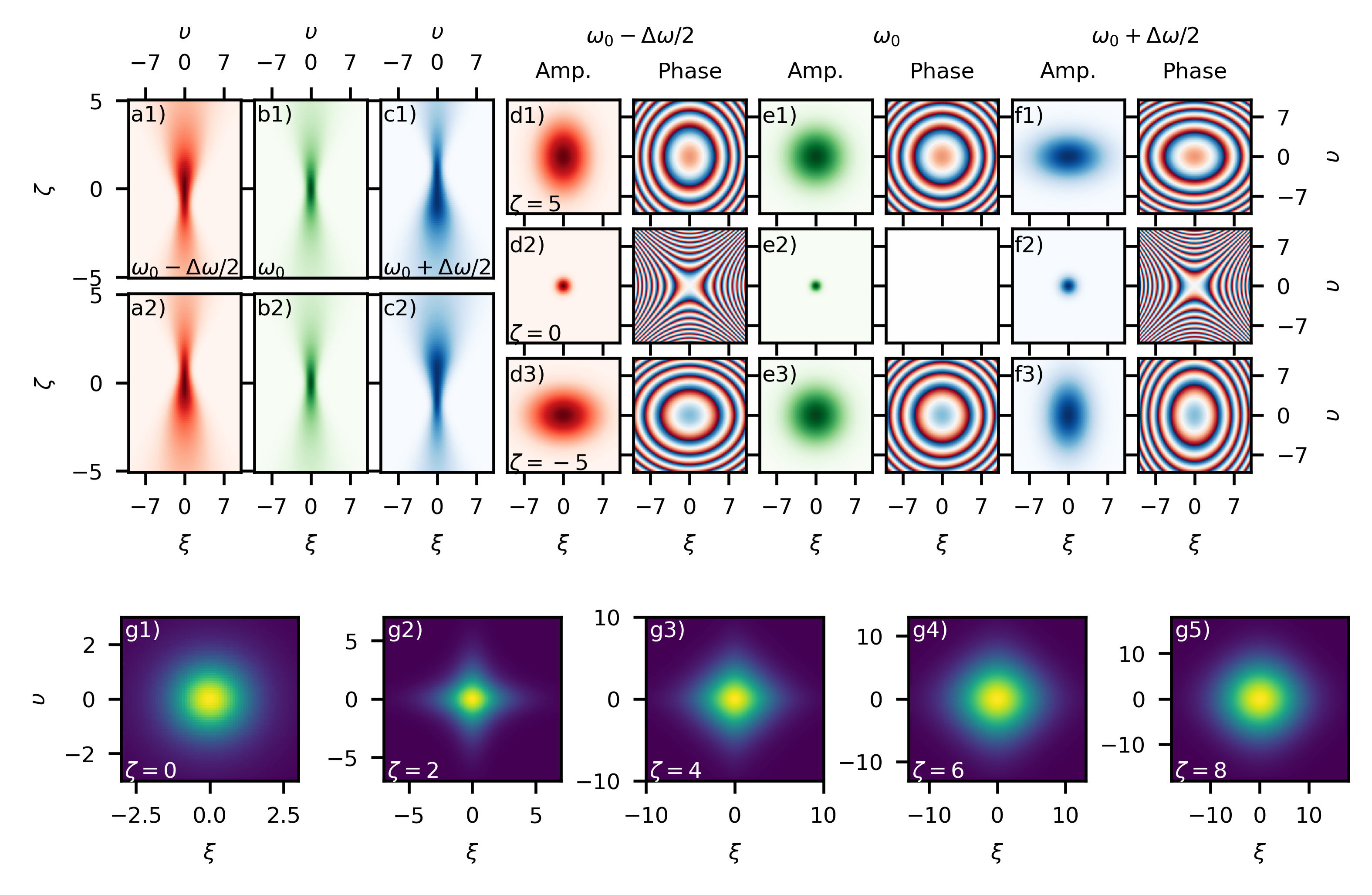}
	\caption{Ultrashort laser pulse with chromatic astigmatism ($\tau_a=\tau_0$) near the focus in frequency space. Slices of the amplitude in $\upsilon-\zeta$ and $\xi-\zeta$ planes at $\xi=0$ and $\upsilon=0$, respectively, are shown for three different frequencies (a--c). Amplitude and phase at slices in $\xi-\upsilon$ planes at three different values of $\zeta$ are shown for the same three frequencies (d--f). The integrated amplitude profile $\int|\tilde{E}|$ is shown (g) for increasing $\zeta$ showing the lack of cylindrical symmetry at intermediate values.}
	\label{fig:FF_freq}
\end{figure*}

Once again the situation is relatively straightforward in frequency space, since each frequency can be considered separate from the rest. For $\tau_p=0$ and $\tau_a=\tau_0$, the results are shown from different perspectives in Fig.~\ref{fig:FF_freq} for $\omega_0$ and $\omega_0\pm\Delta\omega/2$. As a direct result from the opposite sign of the wavefront curvature in the nearfield according to Eq.~\ref{eq:NF_freq}, the different colors now have an asymmetry in their phase and amplitude near the focus.

We can see that the waist positions vary with frequency, and are at opposite $\zeta$ for frequencies on each side of $\omega_0$ and when considering the beam waist either along $\xi$ or $\upsilon$. At $\omega_0$ the beam is round at all $\zeta$ since there is no aberration whatsoever. Frequencies outside of $\omega_0$ are round at $\zeta=0$, but have non-zero and asymmetric spatial phase. Away from $\zeta=0$ all frequencies besides $\omega_0$ are asymmetric in both amplitude and phase. The integrated profile shown in Fig.~\ref{fig:FF_freq}(g) is round when $\zeta=0$ and when $|\zeta|$ becomes large, but at intermediate values there is a clear lack of cylindrical symmetry (but symmetry still when exchanging $\xi$ and $\upsilon$).

\begin{figure*}[tbh]
	\centering
	\includegraphics[width=\linewidth]{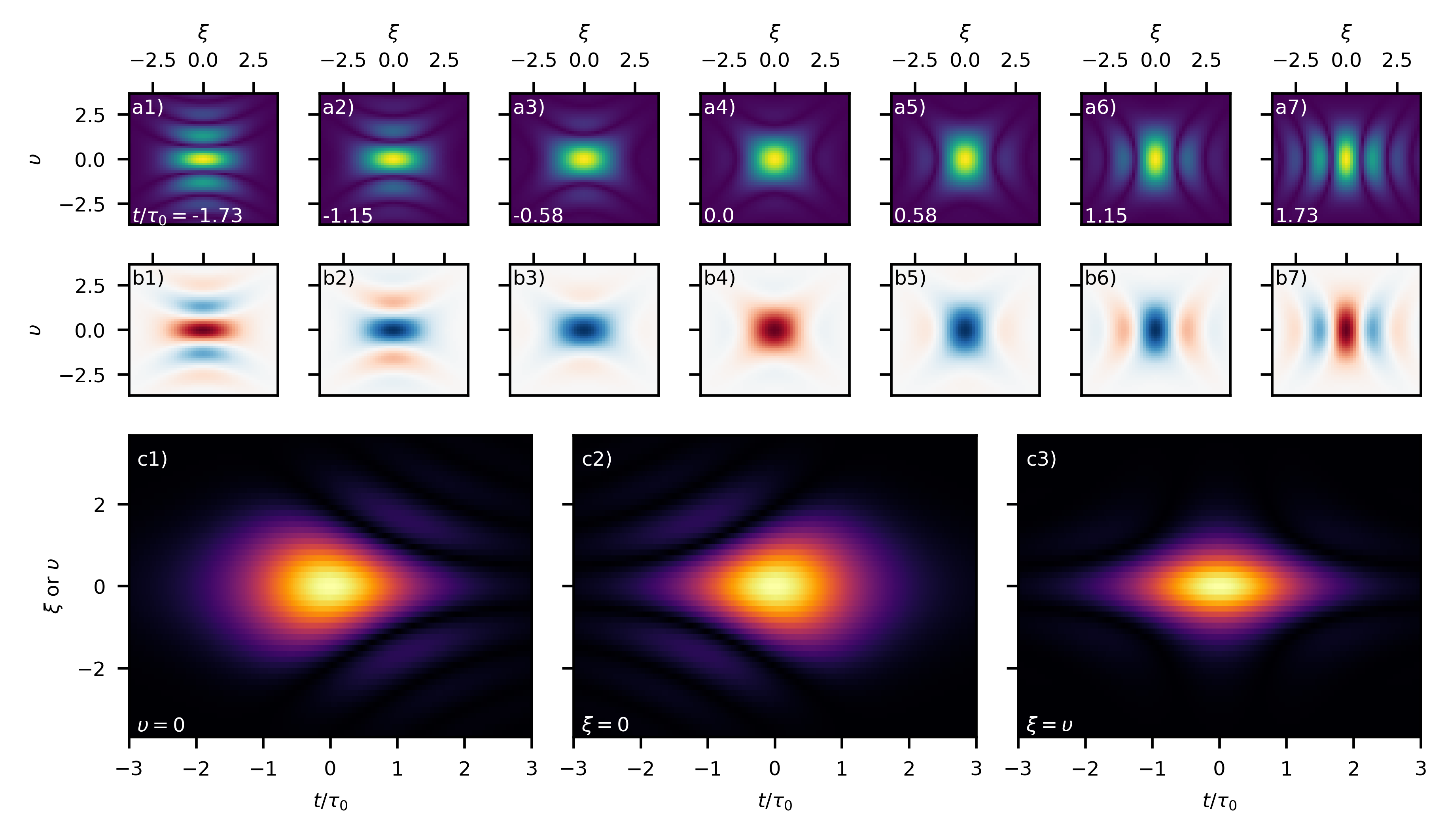}
	\caption{Ultrashort laser pulse with chromatic astigmatism ($\tau_a=\tau_0$) at the focus ($\zeta=0$) in time. The amplitude (a) and real part of the electric field (b) are shown in $\xi-\upsilon$ for 7 different times around $t=0$. The amplitude in space-time (c) for three different transverse slices shows the symmetry expected in space.}
	\label{fig:FF_time}
\end{figure*}

The frequency-dependent fields produced by Eq.~\ref{eq:FF_CAstig} and shown in Fig.~\ref{fig:FF_freq} can be numerically Fourier-transformed to calculate the fields in time. The amplitude and real fields are shown in space for 7 different times in Fig.~\ref{fig:FF_time}(a) and (b), respectively, at the best focus position for $\omega_0$ ($\zeta=0$), along with the space-time amplitude for different spatial slices in Fig.~\ref{fig:FF_time}(h--j).

In the time slices in Fig.~\ref{fig:FF_time}(a) and (b), the symmetries are again observed where a beam at $t'$ is the same as a beam at $-t'$ with $x$ and $y$ exchanged. At $t=0$ the beam profile is symmetric itself with an exchange of $x$ and $y$, but it is markedly not cylindrically symmetric. As $|t|$ increases the amplitude shows spatial interferences in one of the transverse coordinates, producing a smaller central spot and outer fringes, while showing no such interferences in the other transverse coordinate and becoming slightly larger. Since we saw in Fig.~\ref{fig:FF_freq} that all frequencies are symmetric in amplitude at $\zeta=0$ and only larger than at $\omega_0$, these interferences and asymmetries come purely from the chromatically-varying and asymmetric spatial phase. Expanding from Eq.~\ref{eq:FF_CAstig} when $\zeta=0$ we can find the spectral phase of the pulse $\exp{\left(-i\phi(\omega,\xi,\upsilon)\right)}$

\begin{align}
	\phi(\omega,\xi,\upsilon)&= \frac{-1}{2}\left[\arctan{\left(-\tau_x\delta\omega\right)}+\arctan{\left(-\tau_y\delta\omega\right)}\right]-\frac{\xi^2\tau_x\omega\delta\omega/\omega_0}{1+(\tau_x\delta\omega)^2}-\frac{\upsilon^2\tau_y\omega\delta\omega/\omega_0}{1+(\tau_y\delta\omega)^2}\\
	\begin{split}
		&=\left[\tau_p(1-(\xi^2+\upsilon^2))-\tau_a(\xi^2-\upsilon^2)\right]\delta\omega\\
		&+\left[6\xi^2(\tau_p+\tau_a)^3+6\upsilon^2(\tau_p-\tau_a)^3-(2\tau_p^3/3+6\tau_p\tau_a^2)\right]\frac{\delta\omega^3}{6}+\mathcal{O}(\delta\omega^4),
	\end{split}
\end{align}

\noindent where we have ignored a second-order term that is small when the pulse is longer than few-cycle ($\Delta\omega/\omega_0\ll1$). In the case of Fig.~\ref{fig:FF_time} where $\tau_p=0$, the overall time delay due to the term linear in $\delta\omega$ is asymmetric ($\propto\xi^2-\upsilon^2$) along with the third-order phase (term proportional to $\delta\omega^3$). It is instructive to compare to the case where $\tau_a=0$, which is cylindrically symmetric. We will not examine the case where both $\tau_a$ and $\tau_p$ are non-zero, but the spectral phase on-axis and it's dependence on $\zeta$ will be detailed in a later section.

There is a $\pi$ phase shift with each interference, which of course means the real part of the field becomes negative. The slices along $\xi=0$ or $\upsilon=0$ in Fig.~\ref{fig:FF_time}(c1) and (c2) are emblematic of the space-time amplitude of a beam with longitudinal chromatism, except that they are flipped in time, and a slice along $\xi=\upsilon$ in Fig.~\ref{fig:FF_time}(c3) shows clearly the lack of cylindrically symmetry from another viewpoint.

\begin{figure*}[tbh]
	\centering
	\includegraphics[width=100mm]{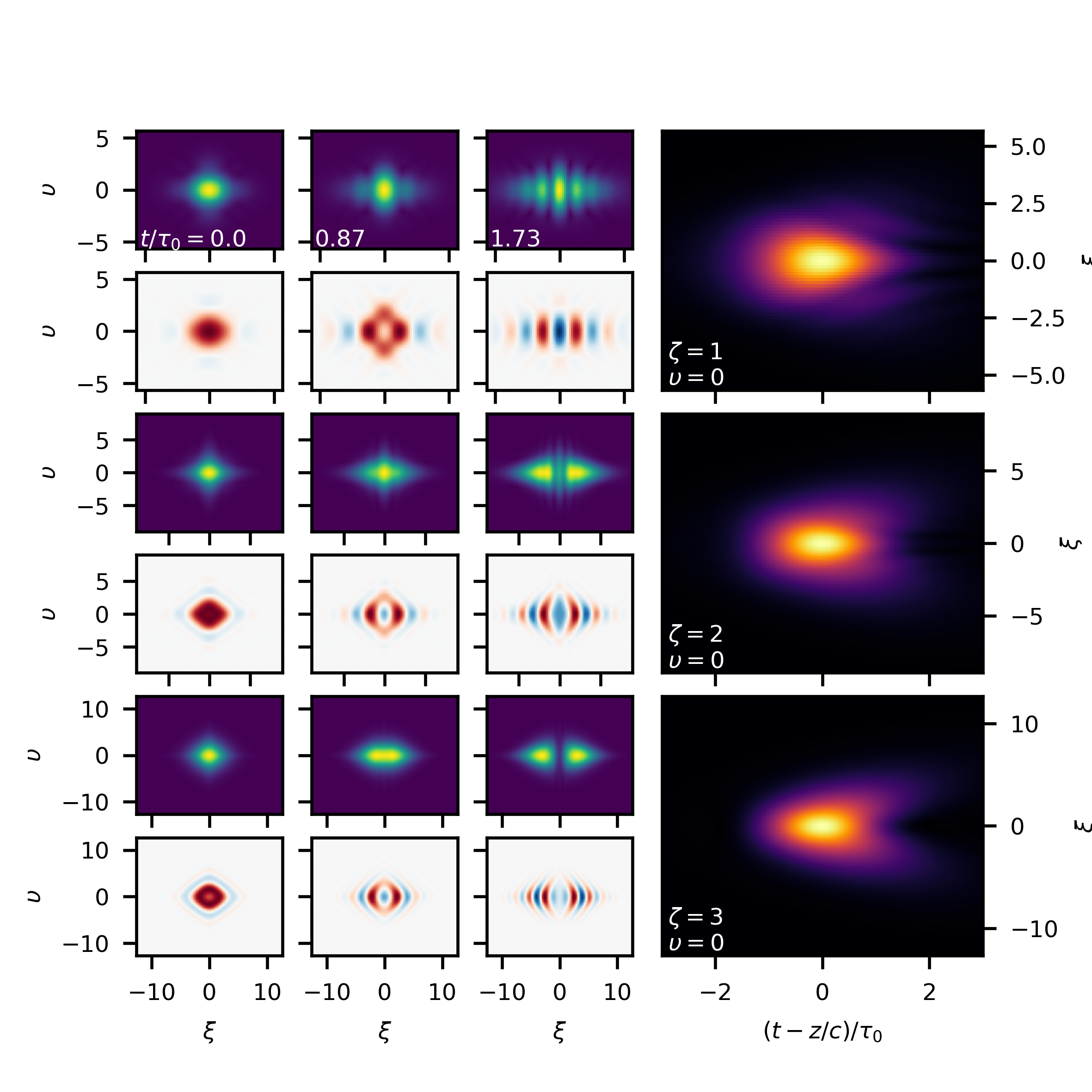}
	\caption{Propagation of an ultrashort laser pulse with chromatic astigmatism ($\tau_a=\tau_0$) near the focus in time. The amplitude and real part of the electric field are shown in $\xi-\upsilon$ planes for three different times (now only positive due to the known symmetry), along with space-time amplitude when $\upsilon=0$, for $\zeta=1$, 2, and 3.}
	\label{fig:FF_time_prop}
\end{figure*}

Since propagation itself can have effects on the space-time distribution of ultrashort fields, the space-time behavior is also interesting away from $\zeta=0$. The amplitude and real fields in space, and the space-time amplitude at $\upsilon=0$ are shown in space for 3 different times (now only positive due to the known symmetry) in Fig.~\ref{fig:FF_time_prop} at $\zeta=$1, 2, and 3. Now that the beam is away from $\zeta=0$ in Fig.~\ref{fig:FF_time_prop}, the asymmetries become more significant. This is due to the fact that, as seen in Fig.~\ref{fig:FF_freq}, both the amplitude and phase are frequency-dependent and asymmetric when $\zeta$ is non-zero.

There is an additional important point on the symmetry around the focus. At $\zeta=0$ as in Fig.~\ref{fig:FF_time} and for the small values of $\zeta$ in Fig.~\ref{fig:FF_time_prop}, the symmetry when reflecting over $t=0$ and exchanging $x$ and $y$ remains. However, when $\zeta$ becomes large, the curvature and resultant time delay become large. Since the curvature is always proportional to $x^2+y^2$ (i.e. not having the same saddle symmetry as the chromatic astigmatism), there is no longer the same symmetry in time.

\section{Analysis with added spectral phase}
\label{sec:chirp}

When second-order spectral phase $\phi_2$ (group delay dispersion, GDD) is added to an ultrashort pulse, the pulse duration generally increases. When the GDD is large enough compared to the Fourier-limited pulse duration ($\phi_2\gg\tau_0$) then it can be approximated that the different frequencies have a linearly varying arrival time as $\phi_2\delta\omega$. It has been shown that combining longitudinal chromatism in the focus of an ultrashort pulse (frequencies focus to different longitudinal positions) and GDD (frequencies arrive at different times) the intensity of the pulse in a region around the focus can travel at velocities much different than the speed of light~\cite{sainte-marie17,palastro18,froula18}. This "flying focus" has been demonstrated with diffractive and refractive optics~\cite{froula18,jolly20-1}, and used to generate ionization waves of tunable velocity~\cite{turnbull18-2}. It is also predicted to influence a number of other experimental scenarios~\cite{ramsey20,ramsey21}.

GDD results in a spectral phase that is quadratic in frequency, taking the form of an additional term $\exp{\left(-i(\phi_2/2)\delta\omega^2\right)}$. When there is LF/PFC ($\tau_p\neq0$) in addition to the significant GDD, then the velocity of the flying focus intensity peak is

\begin{equation}
	v_\textrm{ff}=\frac{c}{1+\frac{c\phi_2}{z_R\tau_p}}.
\end{equation}

\noindent There is a value of GDD $\phi_2^{\textrm{Inf}}=-z_R\tau_p/c$ where the velocity becomes poorly defined and there is in fact no evolution of the focus in space. For positive (negative) $\tau_p$, above $\phi_2^{\textrm{Inf}}$ $v_\textrm{ff}$ is positive (negative) and below it $v_\textrm{ff}$ is negative (positive). When $\phi_2/\tau_p$ is positive then $v_\textrm{ff}<c$, and the converse for when $0>\phi_2/\tau_p>-z_R/c$, and finally when $\phi_2/\tau_p<-z_R/c$ $v_\textrm{ff}$ is both negative and subluminal.

\begin{figure}[tbh]
	\centering
	\includegraphics[width=\linewidth]{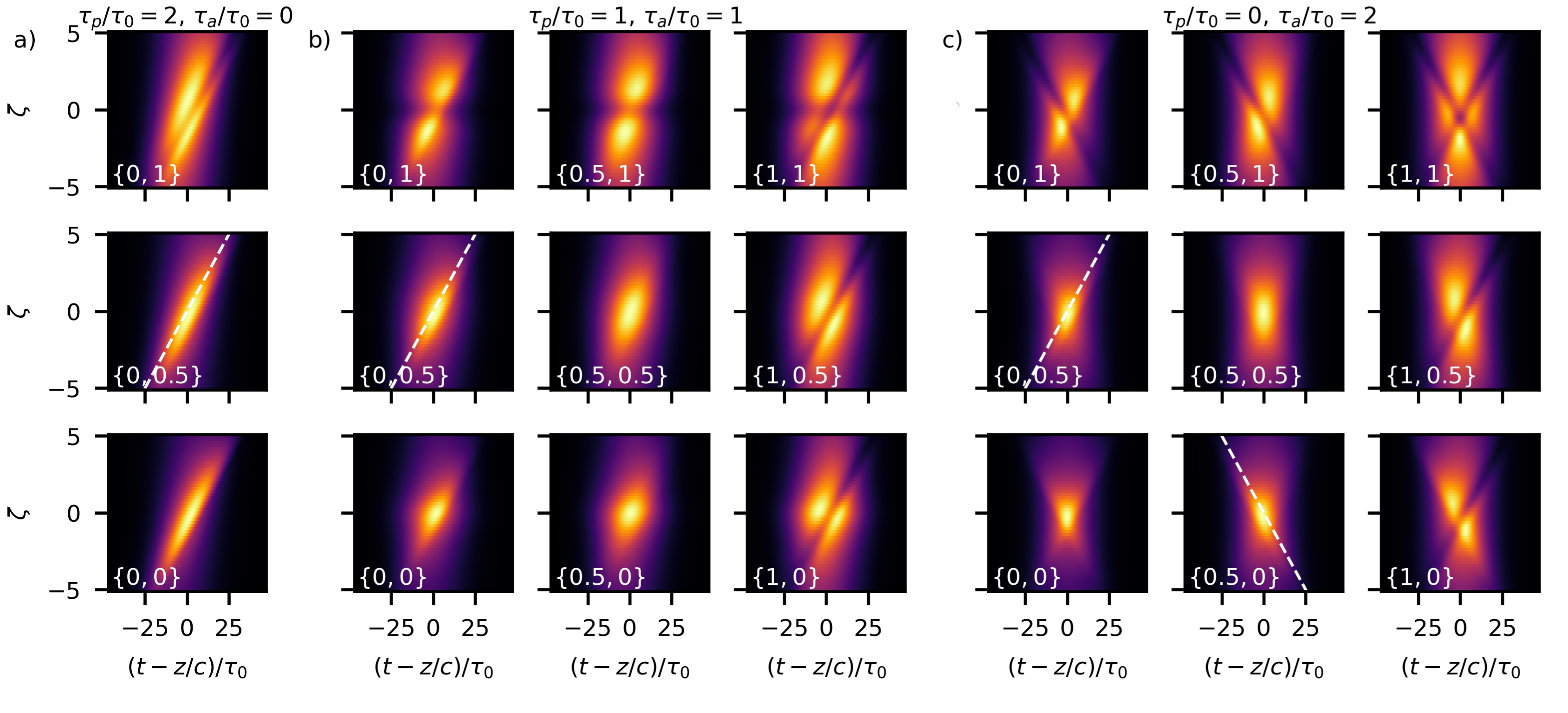}
	\caption{Ultrashort laser pulse with longitudinal chromatism and/or chromatic astigmatism near the focus and additional GDD ($\phi_2=10\tau_0^2$). The pulse has $\tau_p=2\tau_0$ and $\tau_a=0$ in (a), $\tau_p=\tau_a=\tau_0$ in (b), and $\tau_p=0$ and $\tau_a=2\tau_0$ in (c). The brackets in each plot correspond to the \{$\xi,\upsilon$\} values to which they correspond. Note that a structure along a vertical line would be at the speed of light, and the dotted lines correspond to ${v}_\textrm{ff}(\pm\tau_a)$.}
	\label{fig:FF_chirp}
\end{figure}

Due to the fact that chromatic astigmatism is a certain extension of LC/PFC, with asymmetry in $x$ and $y$, there should be some similar space-time effects in the focus of an ultrashort pulse with chromatic astigmatism when GDD is added. An example of these effects can be seen in Fig.~\ref{fig:FF_chirp} with $\phi_2=10\tau_0^2$. A map of the amplitude of an ultrashort pulse is shown for the case of pure LC in Fig.~\ref{fig:FF_chirp}(a) on-axis and at two off-axis positions. There is clear structure that is traveling at a velocity different than $c$, i.e. at $v_\textrm{ff}$ according to $\tau_p=2\tau_0$. Although this effect is less clear off-axis due to interference, it is still present.

We find interesting behavior when the chromatic astigmatism is added, where the intensity of the light pulse has features that travel at different velocities at different points in space. When there is LC and chromatic astigmatism of equal value, shown in Fig.~\ref{fig:FF_chirp}(b), there is effectively only chromatic behavior in $x$ because $\tau_y=0$. when $\upsilon=0$ there is a weak structure traveling at $v_\textrm{ff}$, weaker because the increase of intensity is due to the weaker chromatic behavior only along $x$. This non-luminal structure is slightly more prevalent when $\upsilon\neq0$.

With only chromatic astigmatism, in Fig.~\ref{fig:FF_chirp}(c), there are structures at ${v}_\textrm{ff}(\tau_a)$ and ${v}_\textrm{ff}(-\tau_a)$. When $|x|=|y|$, there is no significant overall structure with a velocity different than $c$, but clear minor structures at  both valid ${v}_\textrm{ff}$ values, since $\tau_x=2\tau_0=-\tau_y$. However, when $|x|\neq|y|$ there is non-luminal structure. For example, when $y=0$ the intensity structure travels at a velocity according to $\tau_x$, and when $x=0$ at a velocity according to $\tau_y$. These observations agree with the symmetries of the equations and the results of the previous section. Indeed, since the effects are mixed when there is chromatic astigmatism, the non-luminal structure is not as clear as with longitudinal chromatism.

\section{Extension to other free-space beams}
\label{sec:extension}

Because the chromatic astigmatism engenders non-trivial spatio-temporal effects in the focus, where the spatial and temporal profiles "mix", ultrashort laser pulses based on different spatial profiles will produce different results. In this section we show the spatio-temporal fields in the focus of an ultrashort Laguerre-Gaussian (LG) beam having chromatic astigmatism. The results on the collimated beam are much more straightforward---simply the higher-order spatial profile with the spatially-varying delay as in Eq.~\ref{eq:NF_time}. Therefore we will focus on the fields around the focus in this section.

The field of a standard Laguerre-Gaussian beam of order $n$ can be written in the compact form as~\cite{kogelnik66,siegman73,jolly22}

\begin{equation}
	\psi^{(\textrm{LG})}=L_n(2ff^*\rho^2)\left(\frac{f}{f^*}\right)^n\psi_0 \label{eq:FF_LG},
\end{equation}

\noindent where $\psi_0$ is from the fundamental Gaussian beam. Just as with the Gaussian beam, the Laguerre-Gaussian beam of order $n$ with chromatic astigmatism can be constructed as follows

\begin{equation}
	\psi_a^{(\textrm{LG})}=L_n\left(2(f_xf_x^*\xi^2+f_yf_y^*\upsilon^2)\right)\left(\frac{f_xf_y}{f_x^*f_y^*}\right)^{n/2}\psi_a,
\end{equation}

\noindent where $\psi_a$ is from the fundamental Gaussian beam with chromatic astigmatism and $f_x$ and $f_y$ are as before, and * denotes the complex conjugate.

\begin{figure*}[tbh]
	\centering
	\includegraphics[width=100mm]{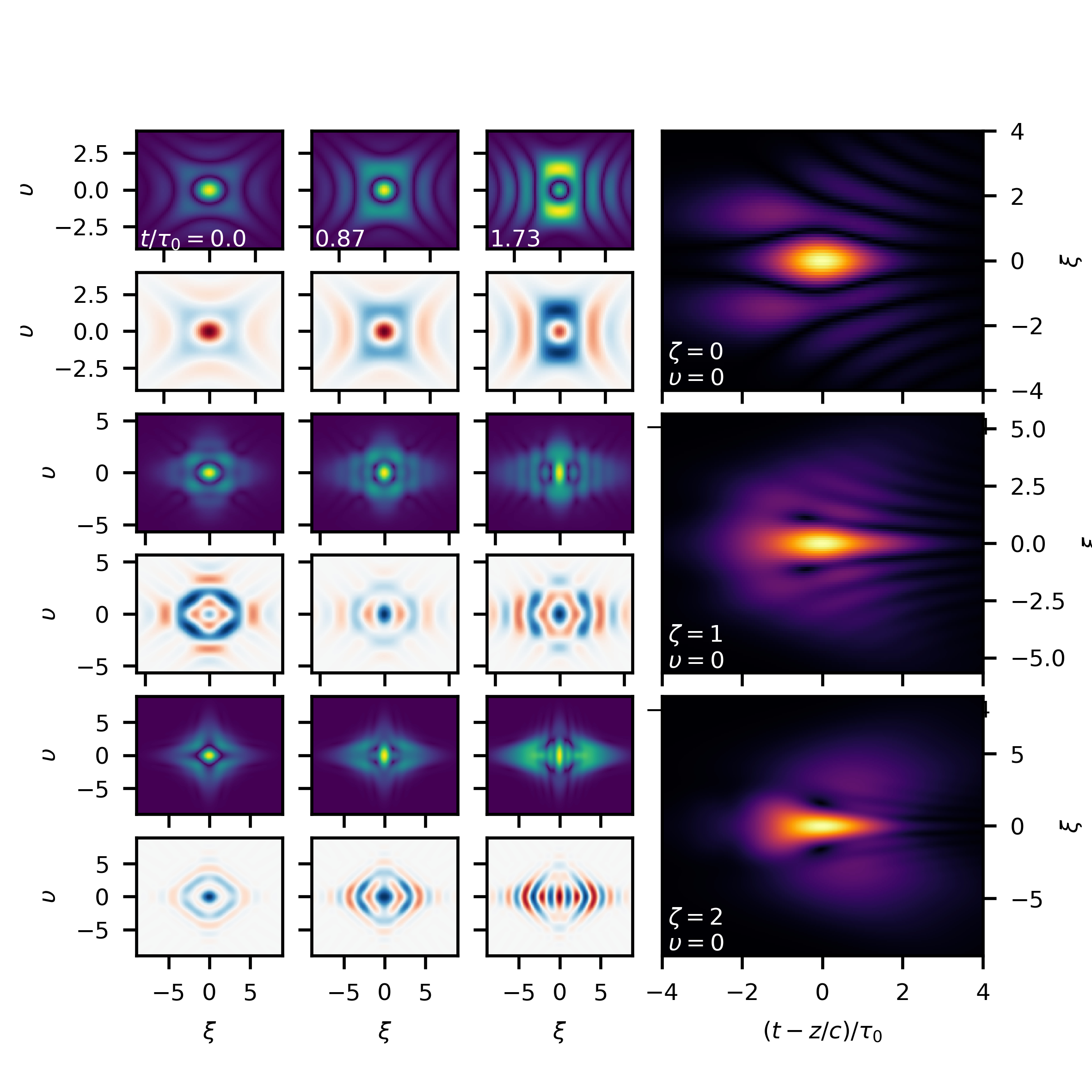}
	\caption{An ultrashort Laguerre-Gaussian ($n=1$) laser pulse with chromatic astigmatism ($\tau_a=\tau_0$) near the focus in time. The amplitude and real part of the electric field are shown in $\xi-\upsilon$ planes for three different times (now only positive due to the known symmetry), along with space-time amplitude when $\upsilon=0$, for $\zeta=0$, 1, and 2.}
	\label{fig:FF_LG}
\end{figure*}

The amplitude and real part of the field for a few time slices and a space-time slice at $\upsilon=0$ are shown in Fig.~\ref{fig:FF_LG} for an $n=1$ LG beam with only chromatic astigmatism ($\tau_a=\tau_0$) for the $\zeta=0$, 1, and 2. There is clearly more space-time complexity, where the interference fringes seen with the fundamental Gaussian are combined with the spatial complexity of the LG beam. A strong asymmetry is also developed as $|\zeta|$ increases, also with more space-time complexity than the fundamental Gaussian case.

\begin{figure*}[tbh]
	\centering
	\includegraphics[width=\linewidth]{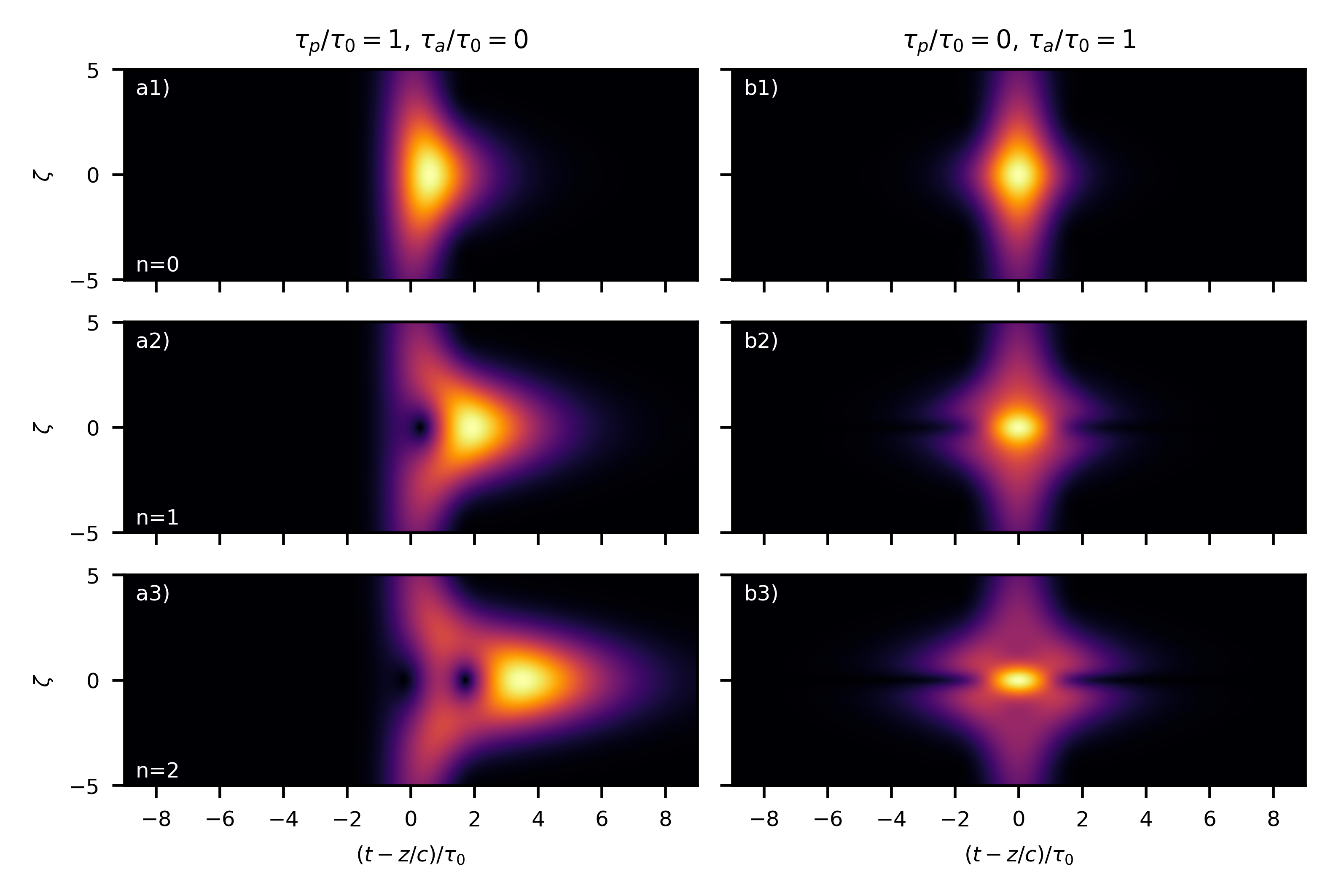}
	\caption{An ultrashort Laguerre-Gaussian laser pulse with pure longitudinal chromatism $\tau_p=\tau_0$ (a) or pure chromatic astigmatism $\tau_a=\tau_0$ (b) near the focus in time, on the optical axis ($\xi=\upsilon=0$).}
	\label{fig:FF_LG_OnAxis}
\end{figure*}

Besides the clearly interesting spatio-temporal profiles presented in Fig.~\ref{fig:FF_LG}, there are also non-trivial effects in the on-axis ($\xi=\upsilon=0$) temporal profiles that warrant a closer look. The on-axis temporal profiles (amplitudes) are shown in Fig.~\ref{fig:FF_LG_OnAxis} over a range of $\zeta$ values for the case of pure longitudinal chromatism (only $\tau_p\neq0$) and pure chromatic astigmatism (only $\tau_a\neq0$) for $n=0$ (Gaussian), 1, and 2.

There are a few defining characteristics of the different cases. The pure LC case, Fig.~\ref{fig:FF_LG_OnAxis}(a), has a time delay near $\zeta=0$ that increases with $n$, where the pure chromatic astigmatism, Fig.~\ref{fig:FF_LG_OnAxis}(b), does not. This makes intuitive sense when considering the nearfield profiles---the pulse-front curvature that produces LC has an overall time delay, where the saddle-shape arrival time has no average arrival time shift. The duration near $\zeta=0$ also increases with $n$ for the pure LC case, where it increases with $n$ near $\zeta=1$ for the chromatic astigmatism case. In both scenarios the duration decreases down to $\tau_0$ as $|\zeta|$ becomes large.

These characteristics can be explained quantitatively when studying the on-axis spectral phase $\phi(\omega,\zeta)$, which is only due to the chromatic nature of the Gouy phase term. This analysis is complementary to that in Section~\ref{sec:analysis_FF} where it was the spatio-spectral phase at $\zeta=0$. According to Eq.~\ref{eq:FF_LG} the field on-axis is proportional to $(f_xf_y/f_x^*f_y^*)^{(n/2)}\sqrt{f_xf_y}$ such that the phase is

\begin{align}
	\phi&= \frac{2n+1}{2}\left[\arctan{\left(\zeta-\tau_x\delta\omega\right)}+\arctan{\left(\zeta-\tau_y\delta\omega\right)}\right]\\
	&=\frac{2n+1}{2}\left[\arctan{\left(\zeta-(\tau_p+\tau_a)\delta\omega\right)}+\arctan{\left(\zeta-(\tau_p-\tau_a)\delta\omega\right)}\right]\\
	&=(2n+1)\arctan{(\zeta)}-\phi(\omega,\zeta),
\end{align}

\noindent such that

\begin{equation}
	\frac{\phi(\omega,\zeta)}{2n+1}=\frac{\tau_p}{1+\zeta^2}\delta\omega+\frac{2\zeta(\tau_p^2+\tau_a^2)}{(1+\zeta^2)^2}\frac{\delta\omega^2}{2}+\frac{2(3\zeta^2-1)(\tau_p^2+3\tau_a^2)\tau_p}{(1+\zeta^2)^3}\frac{\delta\omega^3}{6}+\mathcal{O}(\delta\omega^4).
\end{equation}

If $\tau_a=0$ then the spectral phase is of odd parity at all $\zeta$. The linear term of the spectral phase $\propto(2n+1)\tau_p/(1+\zeta^2)$ and is responsible for the time delay at $\zeta=0$. The third-order term (and to a lesser extent higher-order odd-order terms) is the cause for the increasing duration with $n$ at $\zeta=0$.

If $\tau_p=0$ then the spectral phase only has non-zero even orders, and is zero at $\zeta=0$ (we are reminded now of the spectral and temporal symmetries). Therefore at $\zeta=0$ the duration is independent of $n$ and only depends on the initial spectral bandwidth $\Delta\omega=2/\tau_0$ and the reduction of the bandwidth due to $\tau_a$ (according to the $\sqrt{f_xf_y}$ term). The second-order spectral phase $\propto(2n+1)\zeta\tau_a^2/(1+\zeta^2)^2$ peaks at $\zeta=\pm\sqrt{1/3}$, but since the bandwidth is increasing as $|\zeta|$ increases, the interplay causes the duration to peak at varying $\zeta$ values, larger as $n$ increases.

In all cases the spectral phase goes to zero as $|\zeta|$ becomes large, along with the bandwidth approaching the full. This is why in both Fig.~\ref{fig:FF_LG_OnAxis}(a) and (b) the duration always tends towards $\tau_0$.

\section{Conclusion}
\label{sec:conclusion}

Chromatic astigmatism is a new space-time couplings where the different component frequencies of an ultrashort pulse-beam have different amounts of astigmatism, which we have described in this work theoretically---analytically and using numerical Fourier transformation. Chromatic astigmatism results in a saddle-shaped arrival time on a collimated beam and significantly more complicated behavior around a focus. We used the spectral phase to elucidate effects at and around the focus for both the fundamental Gaussian beam and Laguerre-Gaussian beams.

Chromatic astigmatism has not yet been produced in an experiment, but we believe that a system of chromatic cylindrical lenses could produce chromatic astigmatism in a similar manner to standard chromatic lenses~\cite{jolly20-1}, and we are in the process of demonstrating this. The concept and analysis of chromatic astigmatism presented in this work will be crucial to use and apply this new STC to experiments and applications. As with many other space-time couplings, chromatic astigmatism may find use in ultrafast laser machining, broadband metrology or microscopy, particle manipulation or acceleration, multi-mode photonics, and beyond.

\begin{backmatter}

\bmsection{Funding}
Marie Sklodowska-Curie Actions (801505).

\bmsection{Acknowledgments}
S.W.J. has received funding from the European Union’s Horizon 2020 research and innovation programme under the Marie Skłodowska-Curie grant agreement No 801505.

\bmsection{Disclosures}
The authors declare no conflicts of interest.

\bmsection{Data availability}
Data underlying the results presented in this paper are not publicly available at this time but may be obtained from the authors upon reasonable request.

\end{backmatter}



\begin{thebibliography}{10}
	\newcommand{\enquote}[1]{``#1''}
	
	\bibitem{akturk10}
	S.~Akturk, X.~Gu, P.~Bowlan, and R.~Trebino, \enquote{Spatio-temporal couplings
		in ultrashort laser pulses,} {\protect\JournalTitle{Journal of Optics}}
	\textbf{12}, 093001 (2010).
	
	\bibitem{bourassin-bouchet11}
	C.~Bourassin-Bouchet, M.~Stephens, S.~de~Rossi, F.~Delmotte, and P.~Chavel,
	\enquote{Duration of ultrashort pulses in the presence of spatio-temporal
		coupling,} {\protect\JournalTitle{Optics Express}} \textbf{19}, 17357--17371
	(2011).
	
	\bibitem{debus19}
	A.~Debus, R.~Pausch, A.~Huebl, K.~Steiniger, R.~Widera, T.~E. Cowan,
	U.~Schramm, and M.~Bussmann, \enquote{Circumventing the dephasing and
		depletion limits of laser-wakefield acceleration,}
	{\protect\JournalTitle{Physical Review X}} \textbf{9}, 031044 (2019).
	
	\bibitem{jolly19-1}
	S.~W. Jolly, \enquote{Influence of longitudinal chromatism on vacuum
		acceleration by intense radially polarized laser beams,}
	{\protect\JournalTitle{Optics Letters}} \textbf{44}, 1833--1836 (2019).
	
	\bibitem{palastro20}
	J.~P. Palastro, J.~L. Shaw, P.~Franke, D.~Ramsey, T.~T. Simpson, and D.~H.
	Froula, \enquote{Dephasingless laser wakefield acceleration,}
	{\protect\JournalTitle{Physical Review Letters}} \textbf{124}, 134802 (2020).
	
	\bibitem{caizergues20}
	C.~Caizergues, S.~Smartsev, V.~Malka, and C.~Thaury, \enquote{Phase-locked
		laser-wakefield electron acceleration,} {\protect\JournalTitle{Nature
			Photonics}} \textbf{14}, 475--479 (2020).
	
	\bibitem{jolly20-2}
	S.~W. Jolly, \enquote{On the importance of frequency-dependent beam parameters
		for vacuum acceleration with few-cycle radially-polarized laser beams,}
	{\protect\JournalTitle{Optics Letters}} \textbf{45}, 3865--3868 (2020).
	
	\bibitem{bor93}
	Z.~Bor, B.~R{\'a}cz, G.~Szab{\'o}, M.~Hilbert, and H.~A. Hazim,
	\enquote{Femtosecond pulse front tilt caused by angular dispersion,}
	{\protect\JournalTitle{Optical Engineering}} \textbf{32}, 2501--2504 (1993).
	
	\bibitem{bor88}
	Z.~Bor, \enquote{Distortion of femtosecond laser pulses in lenses and lens
		systems,} {\protect\JournalTitle{Journal of Modern Optics}} \textbf{35},
	1907--1918 (1988).
	
	\bibitem{bor89-1}
	Z.~Bor, \enquote{Distortion of femtosecond laser pulses in lenses,}
	{\protect\JournalTitle{Optics Letters}} \textbf{14}, 119--121 (1989).
	
	\bibitem{alonso18}
	B.~Alonso, J.~P{\'e}rez-Vizcaíno, G.~M{\'i}nguez-Vega, and {\'I}.~J. Sola,
	\enquote{Tailoring the spatio-temporal distribution of diffractive focused
		ultrashort pulses through pulse shaping,} {\protect\JournalTitle{Optics
			Express}} \textbf{26}, 10762--10772 (2018).
	
	\bibitem{sainte-marie17}
	A.~Sainte-Marie, O.~Gobert, and F.~Qu{\'e}r{\'e}, \enquote{Controlling the
		velocity of ultrashort light pulses in vacuum through spatio-temporal
		couplings,} {\protect\JournalTitle{Optica}} \textbf{4}, 1298--1304 (2017).
	
	\bibitem{porras09}
	M.~A. Porras, \enquote{Characterization of the electric field of focused pulsed
		gaussian beams for phase-sensitive interactions with matter,}
	{\protect\JournalTitle{Optics Letters}} \textbf{34}, 1546--1548 (2009).
	
	\bibitem{palastro18}
	J.~P. Palastro, D.~Turnbull, S.-W. Bahk, R.~K. Follett, J.~L. Shaw,
	D.~Haberberger, J.~Bromage, and D.~H. Froula, \enquote{Ionization waves of
		arbitrary velocity driven by a flying focus,} {\protect\JournalTitle{Physical
			Review A}} \textbf{97}, 033835 (2018).
	
	\bibitem{froula18}
	D.~H. Froula, D.~Turnbull, A.~S. Davies, T.~J. Kessler, D.~Haberberger, J.~P.
	Palastro, S.-W. Bahk, I.~A. Begishev, R.~Boni, S.~Bucht, J.~Katz, and J.~L.
	Shaw, \enquote{Spatiotemporal control of laser intensity,}
	{\protect\JournalTitle{Nature Photonics}} \textbf{12}, 262--265 (2018).
	
	\bibitem{jolly20-1}
	S.~W. Jolly, O.~Gobert, A.~Jeandet, and F.~Qu{\'e}r{\'e}, \enquote{Controlling
		the velocity of a femtosecond laser pulse using refractive lenses,}
	{\protect\JournalTitle{Optics Express}} \textbf{28}, 4888--4897 (2020).
	
	\bibitem{turnbull18-2}
	D.~Turnbull, P.~Franke, J.~Katz, J.~P. Palastro, I.~A. Begishev, R.~Boni,
	J.~Bromage, A.~L. Milder, J.~L. Shaw, and D.~H. Froula, \enquote{Ionization
		waves of arbitrary velocity,} {\protect\JournalTitle{Physical Review
			Letters}} \textbf{120}, 225001 (2018).
	
	\bibitem{ramsey20}
	D.~Ramsey, P.~Franke, T.~T. Simpson, D.~H. Froula, and J.~P. Palastro,
	\enquote{Vacuum acceleration of electrons in a dynamic laser pulse,}
	{\protect\JournalTitle{Physical Review E}} \textbf{102}, 043207 (2020).
	
	\bibitem{ramsey21}
	D.~Ramsey, B.~Malaca, A.~D. Piazza, M.~Formanek, P.~Franke, D.~H. Froula,
	M.~Pardal, T.~T. Simpson, J.~Vieira, K.~Weichman, and J.~P. Palastro,
	\enquote{Nonlinear thomson scattering with ponderomotive control,}
	{\protect\JournalTitle{Physical Review E}} \textbf{105}, 065201 (2021).
	
	\bibitem{kogelnik66}
	H.~Kogelnik and T.~Li, \enquote{Laser beams and resonators,}
	{\protect\JournalTitle{Applied Optics}} \textbf{5}, 1550--1567 (1966).
	
	\bibitem{siegman73}
	A.~E. Siegman, \enquote{{Hermite-Gaussian} functions of complex argument as
		optical-beam eigenfunctions,} {\protect\JournalTitle{Journal of the Optical
			Society}} \textbf{63}, 1093--1094 (1973).
	
	\bibitem{jolly22}
	S.~W. Jolly and M.~A. Porras, \enquote{Clarification for the fields of
		different radially polarized laguerre–gaussian light beams,}
	{\protect\JournalTitle{Optics Letters}} \textbf{47}, 3632--3635 (2022).
	
\end{thebibliography}
\end{document}